\documentclass[]{spie}  
\usepackage[]{graphicx}
\usepackage[]{subfigure}
\title{ An X-ray polarimeter for hard X-ray optics}

\author{Fabio  Muleri\supit{a}, Ronaldo Bellazzini\supit{b},
Enrico Costa\supit{a}, Paolo  Soffitta\supit{a}, Francesco Lazzarotto\supit{a},
Marco Feroci\supit{a}, Luigi Pacciani\supit{a}, Alda
Rubini\supit{a}, Ennio Morelli\supit{c}, Luca Baldini\supit{b},
Francesco Bitti\supit{b}, Alessandro Brez\supit{b}, Francesco
Cavalca\supit{b}, Luca Latronico\supit{b},  Marco Maria
Massai\supit{b},  Nicola Omodei$^b$, Michele
Pinchera\supit{b}, Carmelo Sgr{\'o}\supit{b}, Gloria Spandre\supit{b},
Giorgio Matt\supit{d}, Giuseppe Cesare Perola\supit{d}, Oberto
Citterio\supit{e}, Giovanni Pareschi\supit{e}, Vincenzo Cotroneo\supit{e}, Daniele Spiga\supit{e}, Rodolfo Canestrari\supit{e}  \skiplinehalf
\supit{a} Istituto di Astrofisica Spaziale e Fisica Cosmica, Via
del Fosso del Cavaliere 100, I-00133 Roma, Italy;
\\
\supit{b} Istituto Nazionale di Fisica Nucleare, Largo B.
Pontecorvo 3, I-56127 Pisa,  Italy
\\
\supit{c} Istituto di
Astrofisica Spaziale e Fisica Cosmica, Via Gobetti 101, I-40129
Bologna, Italy
\\
\supit{d} Universit{\`a} degli Studi di Roma 3, via
della Vasca Navale 84, I-00146 Roma, Italy
\\
\supit{e} Osservatorio
Astronomico di Brera, via E. Bianchi 46, I-23807 Merate (LC),
Italy }

\authorinfo{Further author information: (Send correspondence to
Fabio Muleri)
\\Fabio Muleri: E-mail: Fabio.Muleri@iasf-roma.inaf.it,
Telephone: +39-0649934565}

\begin{document}
  \maketitle

\begin{abstract}
Development of multi-layer optics makes feasible the use of X-ray telescope at energy up to 60-80~keV: in this paper we discuss the extension of photoelectric polarimeter based on Micro Pattern Gas Chamber to high energy X-rays. We calculated the sensitivity with Neon and Argon based mixtures at high pressure with thick absorption gap: placing the MPGC at focus of a next generation multi-layer optics, galatic and extragalactic X-ray polarimetry can be done up till 30~keV.
\end{abstract}



\section{INTRODUCTION}
\label{sect:intro}  
\subsection{Hard X-Ray Telescopes}
In the History of X-ray Astronomy a crucial turning point was the
introduction of X-ray Optics, first aboard some rockets and
eventually on satellite with the HEAO-2/Einstein mission. Optics play two major
functions:

\begin{itemize}
\item to image a field of X-ray Sky, to study extended sources, to
resolve nearby sources, to localize sources with high angular
accuracy;

\item to increase dramatically the signal to noise ratio, also with respect
to other imaging systems, such as coded masks or modulation
collimators. A source is to be compared with the fluctuations of
background inside the point spread function of the telescope and
and not with the total background of the detector.
\end{itemize}

This tremendous power of X-ray optics has been originally applied
to softer X-rays with Einsten and ROSAT, and has been extended to
higher energies to include the Fe lines region, with ASCA, SAX,
Newton and Chandra. The conventional X-ray optics are limited in
band because of the intrinsic limits set by the laws of total
reflection. The critical angle of reflection is roughly
proportional to the square root of the density of the reflecting
material and to the wavelength of the photons. Thence, as the
margin of increase in density, with respect to gold, is very
limited, the extension to higher energies is committed to inner
shells with progressively smaller projected area. The extension to
higher energies is therefore payed with a large number of shells
and with a proportional reduction of the field of view.
Historically the X-ray telescopes have been used up to 10~keV.
Some extended capabilities of Newton at higher energies have not
been been effective because of an area mismatched with that at
lower energies and because of a high background due to the large
plate-scale.

In the future the revolution of the application of optics is about
to be extended to higher energies, thanks to the use of telescopes
with multi-layer coatings\cite{Pareschi03}, alternating layers of high Z and low Z
materials of gradually changing thickness. This technique
provides effective reflection at angles up to 3 times those of
mono-layer surfaces and makes feasible telescopes effective up to
60-80~keV (in practice up to the K absorption edge of the high Z
reflecting material). At lower energies they are not as good as
conventional optics, due to the photoelectric absorption of the
high Z component, but an improvement could be the recently
proposal of over-coating the multilayer surface with a carbon
layer\cite{Pareschi04}.  New missions based on the concept of multilayer telescopes
have been proposed and NUSTAR and SIMBOL-X have passed a certain
level of selection. Depending on various constructive parameters
these missions can gain from two to three orders of magnitude in
sensitivity with respect to the best previous experiments based on
collimation (PDS) or coded mask (IBIS). Also future big telescopes
for X-ray Astronomy, such as NEXT, CON-X or XEUS should include
such optics. The science to be performed with these telescopes is
mainly based on spectroscopy (cyclotron lines, hard tails,
Compton-thick objects etc.). Also imaging can play a role. E.g.
the Diffuse X-ray Background peaks around 40~keV. A hard X-ray
telescope could resolve the large majority of it into discrete
sources. Moreover imaging of extended objects with hard
non-thermal components, such as some Super Nova Remnants and some
Galaxy Clusters, could help to localize the hard emission and
correlate it to the dynamics of the system.

\subsection{X-Ray Polarimetry}
In this paper we want to discuss another point: the capability to
perform Polarimetry of X-ray Sources in the Hard X-ray Band and
the benefit to employ multi-layer optics for this purpose.

Since the birth of X-ray Astronomy, theoretical analysis has
suggested that measuring the linear polarization of X-rays could
significantly improve our knowledge of Physical Processes and of
geometry of regions of the source emitting X-rays and its transfer 
toward the observer.

Here the question rises whether the extension to higher energies
is only a matter of completeness or, on the contrary, a specific
science is there that can only be performed at best at higher
energies. We try to lay down a few hot topics of High Energy
Astrophysics, for which the polarimetry at higher energies would
be more easy or more significant.
\begin{itemize}
\item X-ray pulsator, resonance frequency, inversion of the
polarization plane (normal - anomalous), higher polarization;

\item polarization by scattering on disks: LMXRB, QSO, Seyfert-1;

\item non thermal components in extended objects: SNR, Clusters
out of the thermal component;
\item effects of GR on disks: Cyg X-1, AGN (increasing effects
closer to the inner orbits);
\item blazars: transition from synchrotron to inverse Compton.
\end{itemize}

From these items it is apparent the strong interest to extend the
measurements of polarization to hard X-rays. Is this feasible?

The conventional approach to X-ray polarimetry includes two major
techniques:

\begin{itemize}
\item Bragg Diffraction at $45^\circ$. It provides an excellent
response to the polarization, conventionally expressed with a
modulation factor $\mu$ close to 1. Conversely the overall
efficiency of the process is very poor because only the photons
within a narrow band around the angle that satisfies the Bragg
condition are reflected. The efficiency can be increased by the use of mosaic
crystals providing a larger band-pass.

\item Compton scattering. It provides a good analyzer of the linear
polarization if the system is designed in such a way to limit the
scattering angles  around $90^\circ$. Since the scattered photons
will be distributed in angle according to the Klein-Nishina cross
section, this will automatically limit the efficiency of the
instrument. The optimal sensitivity will derive from a trade-off
on the acceptance angles.
At low energies, when the energy lost in the scattering is low,
the scatterer must be a passive 
low Z material, usually Lithium, encased in a thin beryllium
can, to give mechanical consistency and to prevent oxidation or
nitridation. This device loses the imaging properties of the
optics and is background dominated.
If the scatterer is not geometrically perfect and if the
pointing does not keep the beam exactly on the axis of the
scatterer, huge spurious polarizations are generated. At higher
energies an active scatterer can be used, such as a detector based on
plastic scintillator or, at even more high energies, on silicon.
Such a device could be in principle position sensitive. But the
active scattering polarimeters built so far never arrived below
50~keV. Active scatterers based on Silicon will be likely operative at
energies even higher.
\end{itemize}

Bragg diffraction allowed the first and till now unique measure of X-ray polarization from an astronomical source: the polarimeter on board OSO-8 satellite, observing Crab Nebula for 3 days, succeeded to detect a polarization of 19.22$\pm$0.92\% at 2.6 and 5.2~keV aligned with optical polarization\cite{Weisskopf76}. This confirmed that X-ray emission is due to synchrotron process, the same as other wavelengths.

Bragg diffraction and Compton scattering, in a focal plane version, were employed at
best in the Stellar X-Ray Polarimeter\cite{Kaaret89}. A thin pyrolithic graphite
crystal was set above the focal plane to reflect
(diffract) photons and to convey them to a detector on a
\textit{secondary} focal plane. The rotation of the whole would
modulate the diffracted flux according to the amount and angle of
polarization of the photons. The higher energy photons would pass
the graphite, impinge on a lithium stick. Four detectors all
around in a well configuration would detect scattered photons.
SXRP was supposed to fly aboard the SRG Mission, that was never
completed. But notwithstanding the huge SODART optics it would be
dominated by the background at source level of a fraction of Crab.

\section{The Micropattern Detector}

The relatively recent discovery of device capable to measure the
linear polarization of X-rays by means of the angular
distribution of photoelectrons provides an alternative.
The Micropattern Gas Chamber\cite{Costa01}$^,$\cite{Bellazzini06a}$^,$\cite{Bellazzini06b} has excellent imaging properties, is
relatively broad band with an energy resolution suitable for
continua and photons can be tagged with time at $\mu$sec level.

The recent development of ASIC chips including anode pads and a
complete Front End Electronics make this device very compact and
easy. The MPGC is discussed in other papers in this same book (see ref.~\citenum{Bellazzini06a} and ref.~\citenum{Bellazzini06b}).
We discuss here how and how much the technique can be applied in
the Hard X-Ray range, covered by multi-layer optics. Our considerations are mainly based on computations and
simulations.

The major
adaptation of the MPGC for applications on Hard X-Rays is the
change of the gas mixture in composition and pressure and the
change in the thickness of the absorption/drift gap. We stress the
point that the simulations performed here are based on the same
programs we use at lower energies. This extension is very reliable
because we are using equations and tabulations in a range where
they are even better determined.

\begin{figure}
\begin{center}
\begin{tabular}{c}
\includegraphics[angle=90,height=7cm]{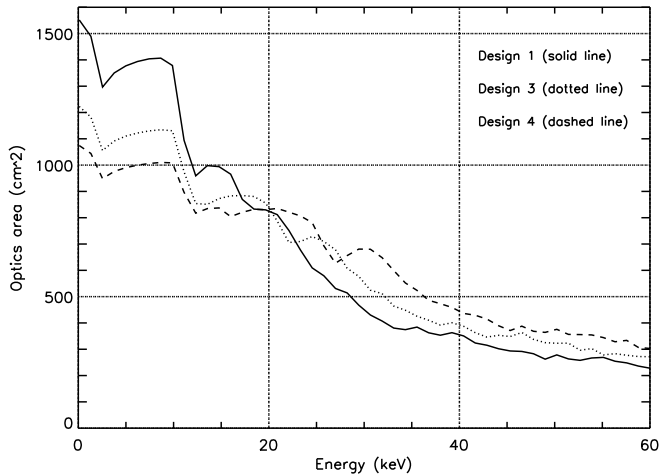}
\end{tabular}
\end{center}
\caption[example]{Area vs energy dependence of three different SIMBOL-X design.}
\label{fig:simbolXArea}
\end{figure}

We discuss the performance and the optimization of our detector
combining its response with that of the optics. As
a bench-mark optics for our simulations we use three of the various
designs of SIMBOL-X mission, described in tab.\ref{tab:simbolXData}. In fig.\ref{fig:simbolXArea} we
indicate the effective area of the telescope alone.

\begin{table}[ht]
\caption{Main features of the telescopes used as a bench-mark. }
\label{tab:fonts}
\begin{center}
\begin{tabular}{|l|c|c|c|} 
\hline
\rule[-1ex]{0pt}{3.5ex}                    &  Design 1  &  Design 3  &  Design 4 \\
\hline
\rule[-1ex]{0pt}{3.5ex}  Focal lenght (m)  &  20        &  25        &  30       \\
\hline
\rule[-1ex]{0pt}{3.5ex}  Max diameter (cm) &  70        &  70        &  70       \\
\hline
\rule[-1ex]{0pt}{3.5ex}  Weight (kg)       &  215       &  229       &  239      \\
\hline
\rule[-1ex]{0pt}{3.5ex}  FoV (arcmin)      &  12.7      &  10.9      &  9.2      \\
\hline
\rule[-1ex]{0pt}{3.5ex}  FoV area (arcmin$^2$) &  161.6 &  118.5     &  85.2     \\
\hline
\end{tabular}
\end{center}
\label{tab:simbolXData}
\end{table}

As to the detector we assume the last generation of the ASIC chip
of 105k pixel at 50$\mu$m pitch (see ref.~\citenum{Bellazzini06a} and ref.~\citenum{Bellazzini06b}) and let free
to vary the thickness of the absorption gap and mixture pressure. We considered Neon and Argon based mixtures with a small percentage of DME (20\% and 40\% respectively).

As to the filling mixture we can fix some general principles:

\begin{itemize}
\item To increase the efficiency $\epsilon$ and extending at higher energy MPGC technique, filling gas of
higher atomic number or higher pressure should be used;
\item Modulation $\mu$ is larger for low pressure or low Z mixture;
\item Increasing thickness of the absorption gap implies an efficiency increase but a modulation decrease.
\end{itemize}

Since sensitivity, expressed as a Minimum Detectable Polarization ({\it MDP}), is inversely
proportional to $\mu \sqrt{\epsilon~A}$ where $A$ is optics area, we need a
trade-off between efficiency and modulation in telescope band-pass.



\subsection{Sensitivity calculation}

Modulation factor for each mixture was calculated from simulating photoelectric absorption, photoelectron propagation and charge collection in the mixture\cite{Bellazzini03}. Absorption from K and L shells was considered, assuming for energy larger than K edge the subshells cross section ratio and 2-p asymmetry parameter at K edge. In tab.~\ref{tab:SubShellCrossSectionRatio} we report subshell emission probability and the 2-p asymmetry parameter at K edge for some mixture elements: only a small fraction of photons are absorbed by 2-p subshell which is not completely modulated.

\begin{table}[ht]
\caption{Subshell emission probability and the 2-p asymmetry parameter for some mixture elements \cite{Yeh93}.}
\label{tab:SubShellCrossSectionRatio}
\begin{center}
\begin{tabular}{|l|c|c|c|c|c|} 
\hline
\rule[-1ex]{0pt}{3.5ex}  Element & K edge (keV)  & $P_{1s}$  & $P_{2s}$ & $P_{2p}$ & Asymmetry parameter\\
\hline
\rule[-1ex]{0pt}{3.5ex}  C       &   0.284  & 95.2\% &  4.2\%   &  0.6\%  & 0.958\\
\hline
\rule[-1ex]{0pt}{3.5ex}  O       &  0.543   & 94.1\% &  4.3\%   &  1.6\%  & 0.955\\
\hline
\rule[-1ex]{0pt}{3.5ex}  Ne      &  0.870   & 93.3\% &  4.2\%   &  2.5\%  & 0.955 \\
\hline
\rule[-1ex]{0pt}{3.5ex}  Ar      &  3.206   & 92.0\% &  5.1\%   &  2.9\%  & 0.820\\
\hline
\end{tabular}
\end{center}
\end{table}

We calculated modulation factor at a given energy simulating absorption of 25000 photons which are directed orthogonally respect to ASIC chip. The optics focusing on detector skew direction of photons up till angle $\sim3^\circ$: simulations suggest that the only effect is a slightly decreasing of the modulation factor.

\begin{figure}
\begin{center}
\begin{tabular}{c}
\subfigure[\label{fig:RangeNeDME8020}]{\includegraphics[angle=0,width=8.2cm]{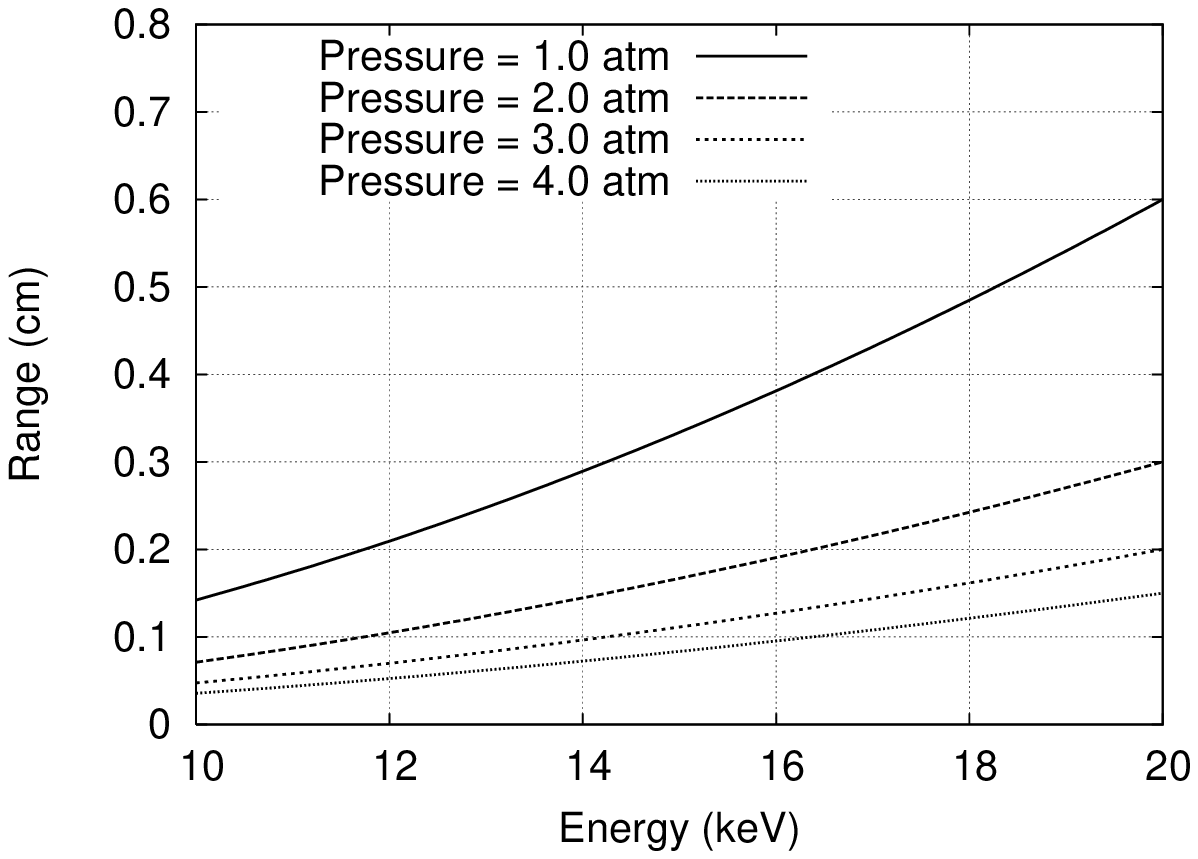}}
\subfigure[\label{fig:RangeArDME6040}]{\includegraphics[angle=0,width=8.2cm]{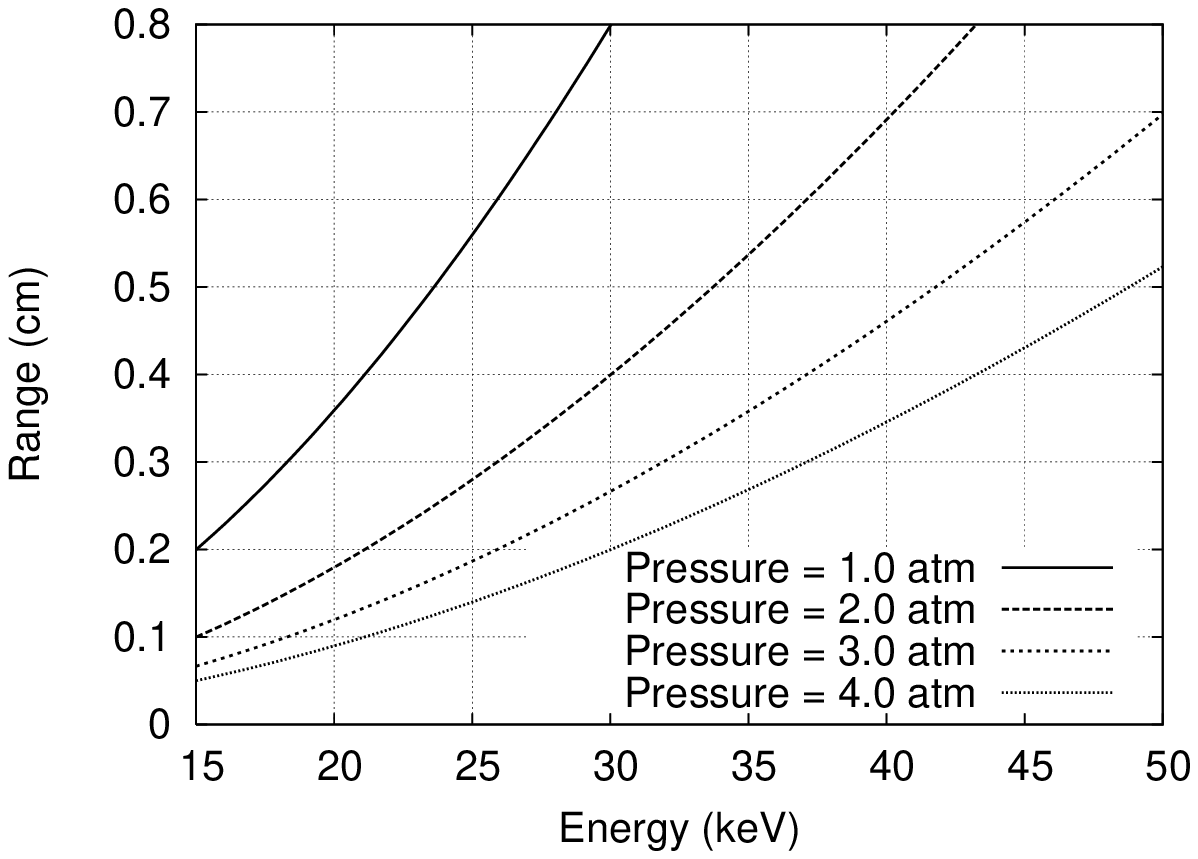}}
\end{tabular}
\end{center}
\caption[example]{Approximate electron range dependence for Ne~80\%~+~DME~20\% ({\bf a}) and Ar~60\%~+~DME~40\% mixture ({\bf b}).}
\end{figure}

Only photoelectrons which were completely absorbed in mixture are considered: so the energy range is upper limited such that the photoelectron range is less than the chip dimensions (about 15x15~mm$^2$). In fig.~\ref{fig:RangeNeDME8020} and fig.~\ref{fig:RangeArDME6040} we report the approximate electron range dependence for Ne and Ar based mixtures at different pressure to give a raw upper limit to the MPGC extension at high energy.

The energy lower limit for Ne and Ar based mixtures was about twice K shell edge because photoelectron emission is often followed by Auger electron production: if its energy is larger than that of the photoelectron, standard reconstruction algorithm may fail in track analysis. In tab.~\ref{tab:FluorescenceYield} we report K edge energy and fluorescence yield for some mixture elements.

The MPGC energy upper limit can be overcome by using mixtures based on higher Z element, such as Krypton or Xenon: in this case the high probability of fluorescence produces a certain number of combinations of low energy electrons that does not allow the straightforward approach we followed so far. We are still investigating this issue.

\begin{table}[ht]
\caption{K shell edge and fluorescence yield for some elements \cite{Krause79}.}
\label{tab:FluorescenceYield}
\begin{center}
\begin{tabular}{|l|c|c|c|} 
\hline
\rule[-1ex]{0pt}{3.5ex}  Element & K edge (keV)  & Fluorescence yield K & Fluorescence yield L\\
\hline
\rule[-1ex]{0pt}{3.5ex}  C       &  0.284        & 0.26\%               &  negligible         \\
\hline
\rule[-1ex]{0pt}{3.5ex}  O       &  0.543        & 0.69\%               &  negligible         \\
\hline
\rule[-1ex]{0pt}{3.5ex}  Ne      &  0.870        & 1.50\%               &  negligible         \\
\hline
\rule[-1ex]{0pt}{3.5ex}  Ar      &  3.206        & 12.0\%               &  $\sim$0.02\%      \\
\hline
\rule[-1ex]{0pt}{3.5ex}  Kr      &  14.33        & 65.2\%               &  $\sim$2.0\%        \\
\hline
\rule[-1ex]{0pt}{3.5ex}  Xe      &  34.59        & 88.8\%               &  $\sim$8.0\%         \\
\hline
\end{tabular}
\end{center}
\end{table}

All the set of data are analysed in sensitivity calculation: however we alredy know that cuts on track shape would increase modulation factor. This would imply an efficiency decrease and so the cuts must be fitted very well with every mixture to ensure the best MDP.

\subsubsection{High Pressure Neon Mixtures}
\label{sect:Neon}

A baseline solution is to use mixtures based on high pressure Ne with thick absorption gap. Computations involve mixtures of 80\% of Neon and DME at 1~atm, 2~atm 3~atm and 4~atm pressure and 1~cm, 2~cm, 3~cm absorption gap thickness.

\begin{figure}
\begin{center}
\begin{tabular}{c}
\subfigure[\label{fig:Modulation_NeDME8020_1cm}]{\includegraphics[angle=90,width=7.35cm]{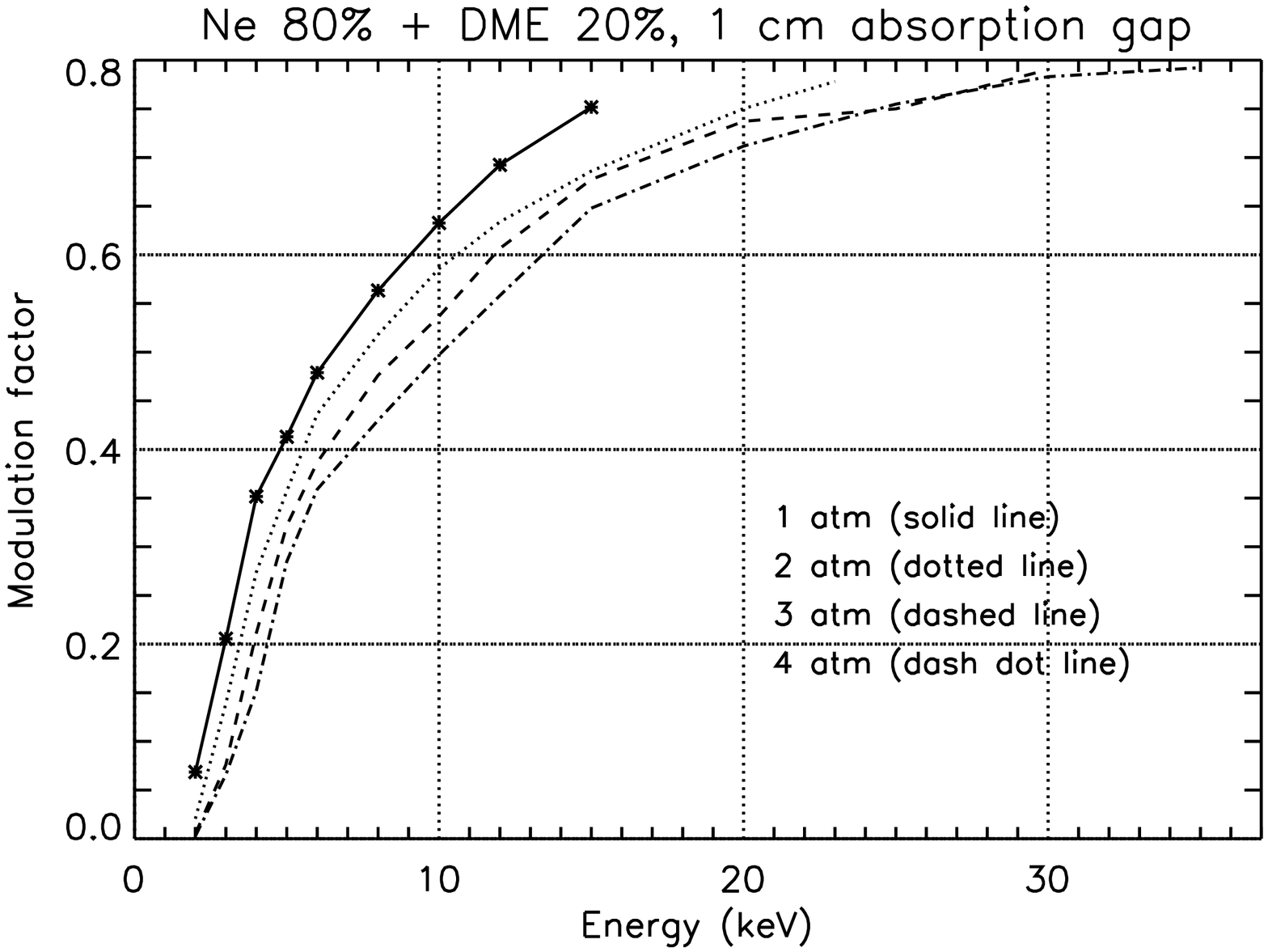}}
\subfigure[\label{fig:MuSqrtE_NeDME8020_1cm}]{\includegraphics[angle=90,width=7.35cm]{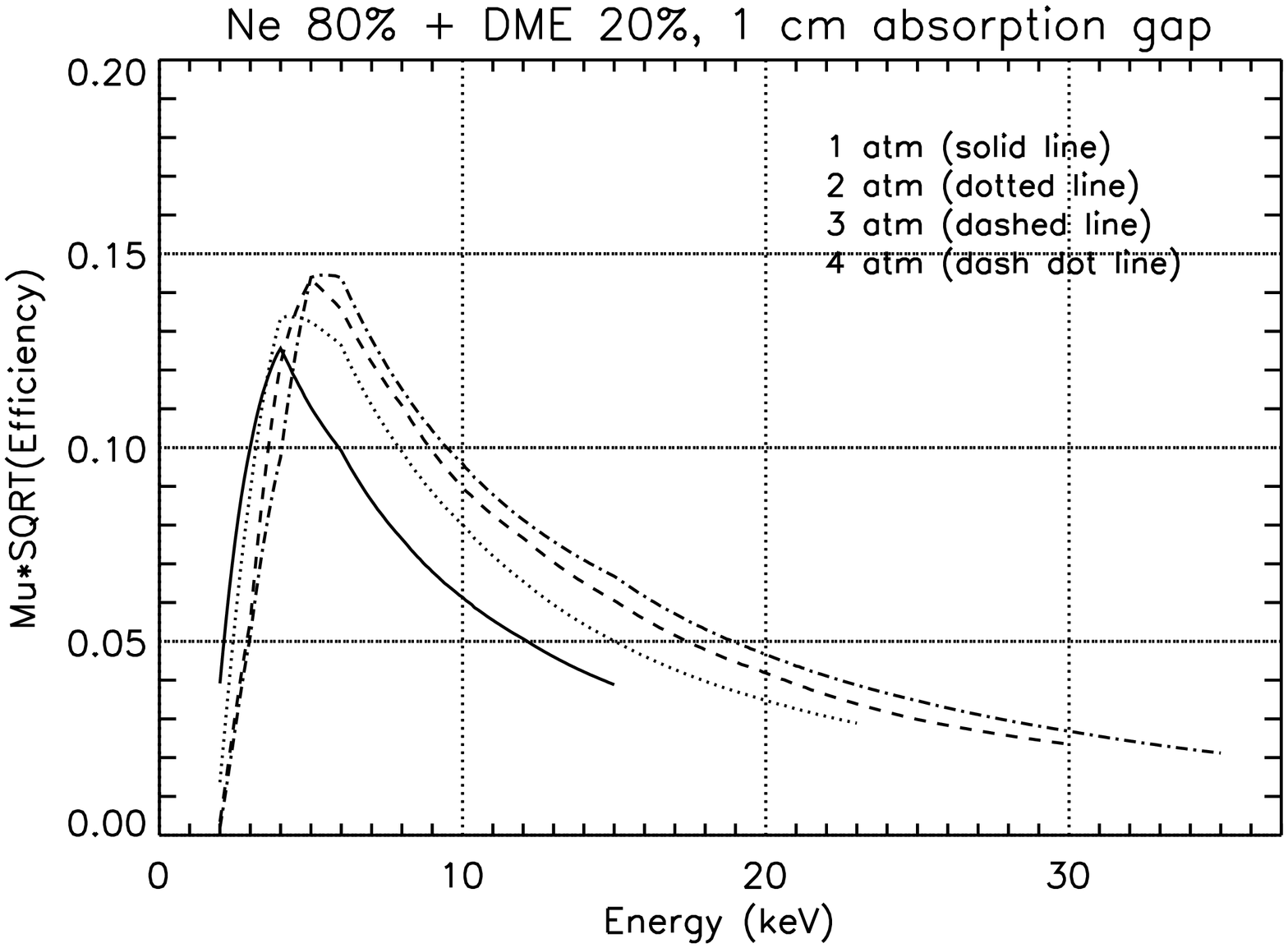}}
\\
\subfigure[\label{fig:Modulation_NeDME8020_2cm}]{\includegraphics[angle=90,width=7.35cm]{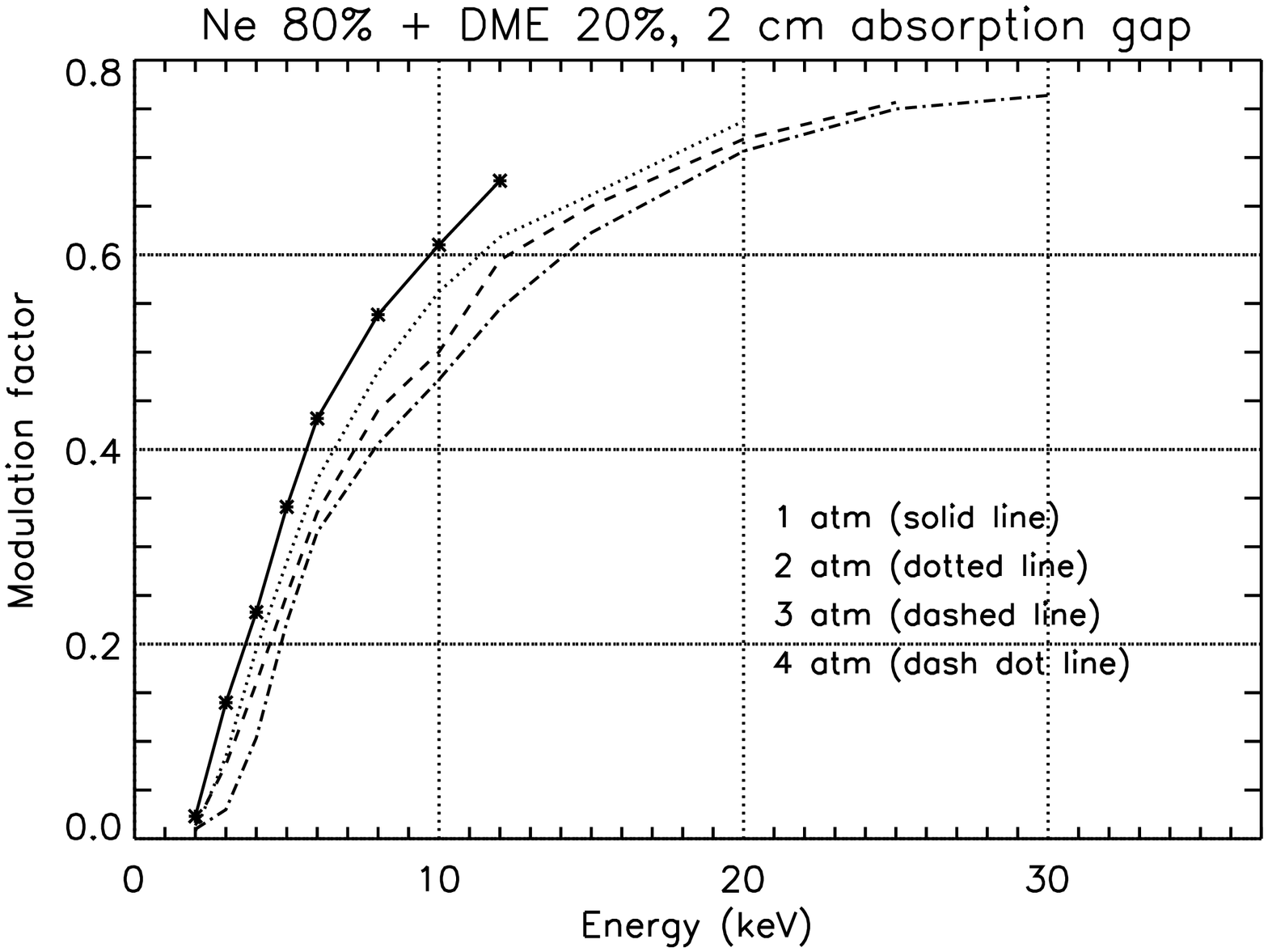}}
\subfigure[\label{fig:MuSqrtE_NeDME8020_2cm}]{\includegraphics[angle=90,width=7.35cm]{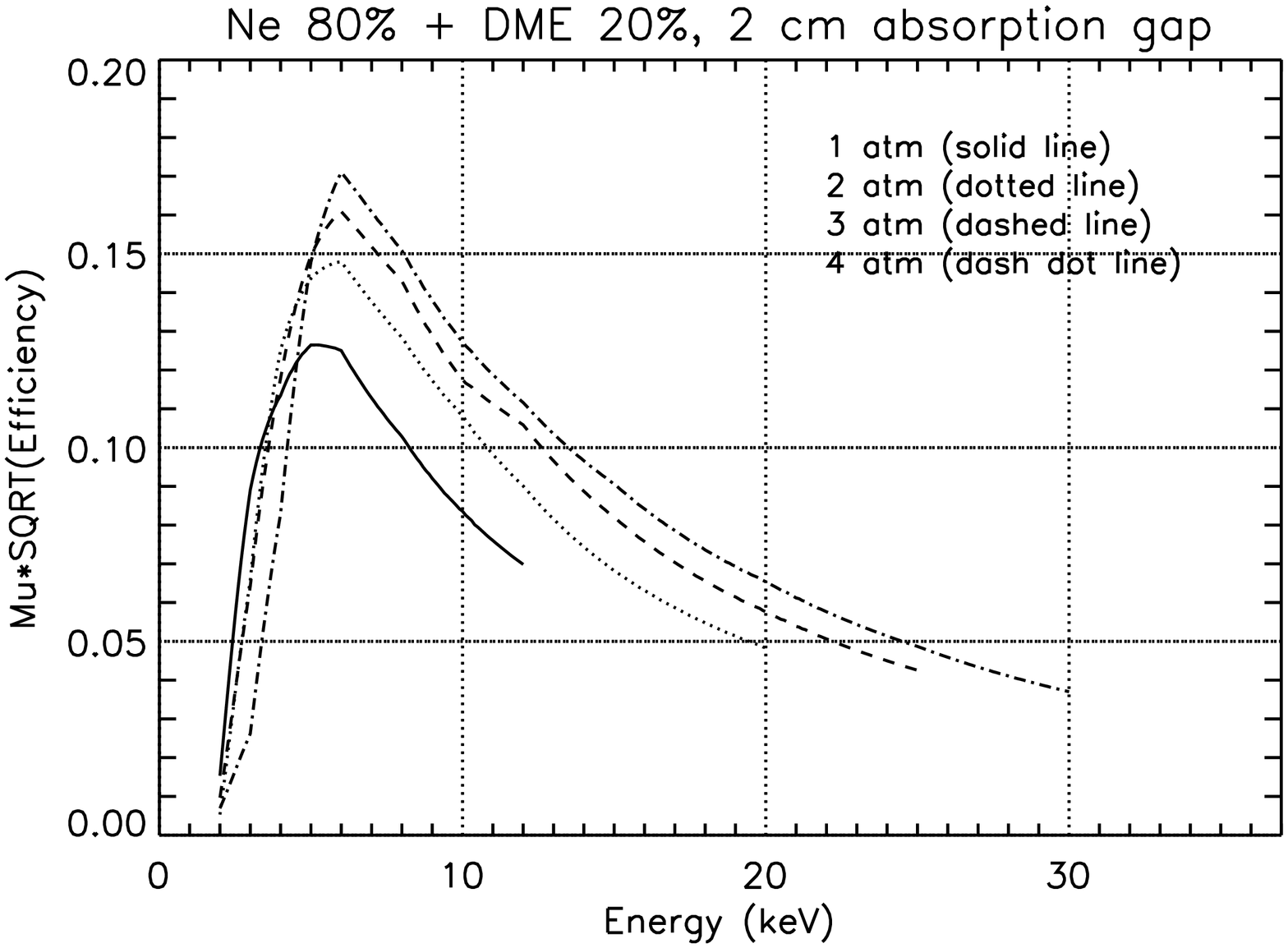}}
\\
\subfigure[\label{fig:Modulation_NeDME8020_3cm}]{\includegraphics[angle=90,width=7.35cm]{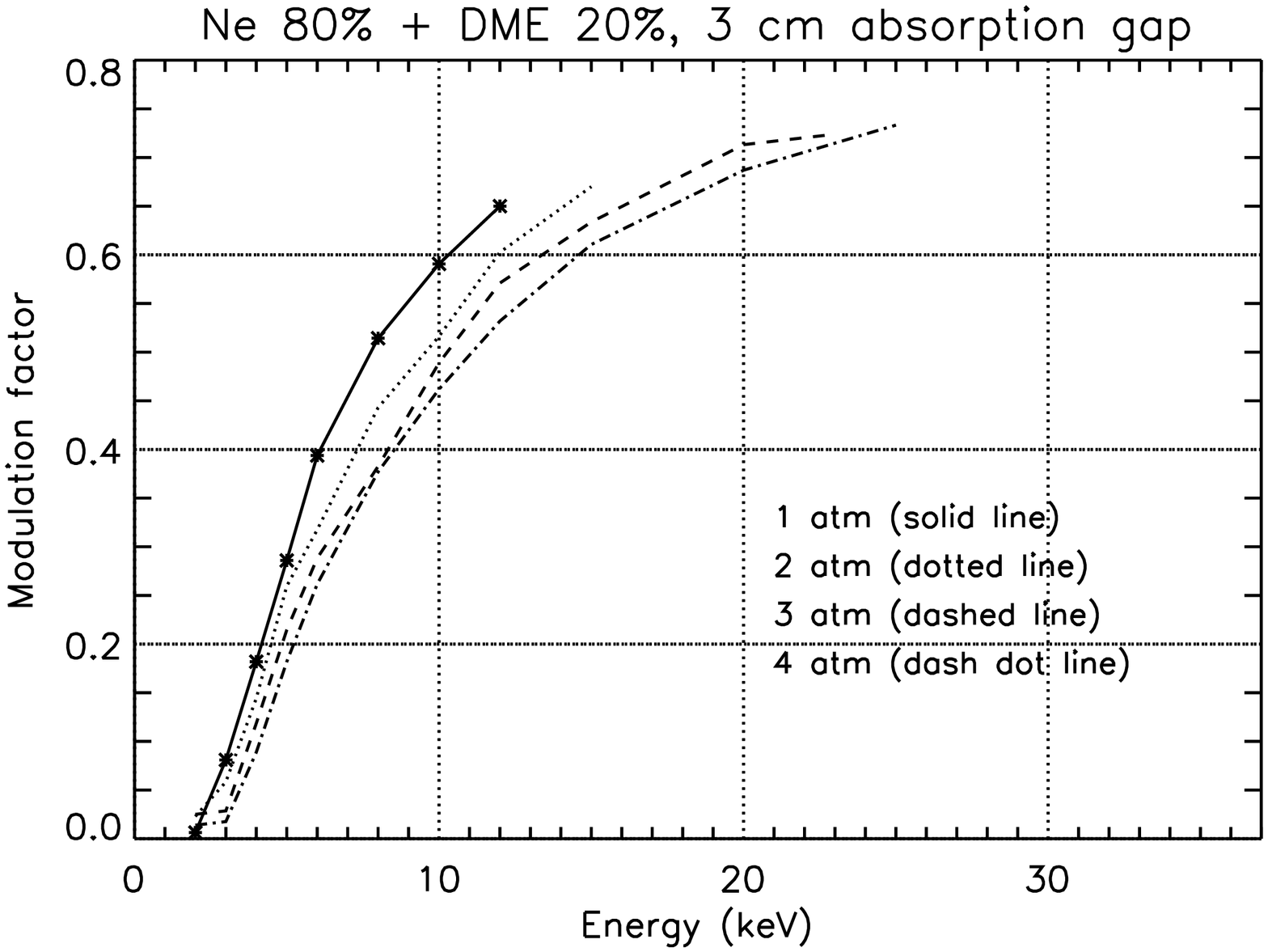}}
\subfigure[\label{fig:MuSqrtE_NeDME8020_3cm}]{\includegraphics[angle=90,width=7.35cm]{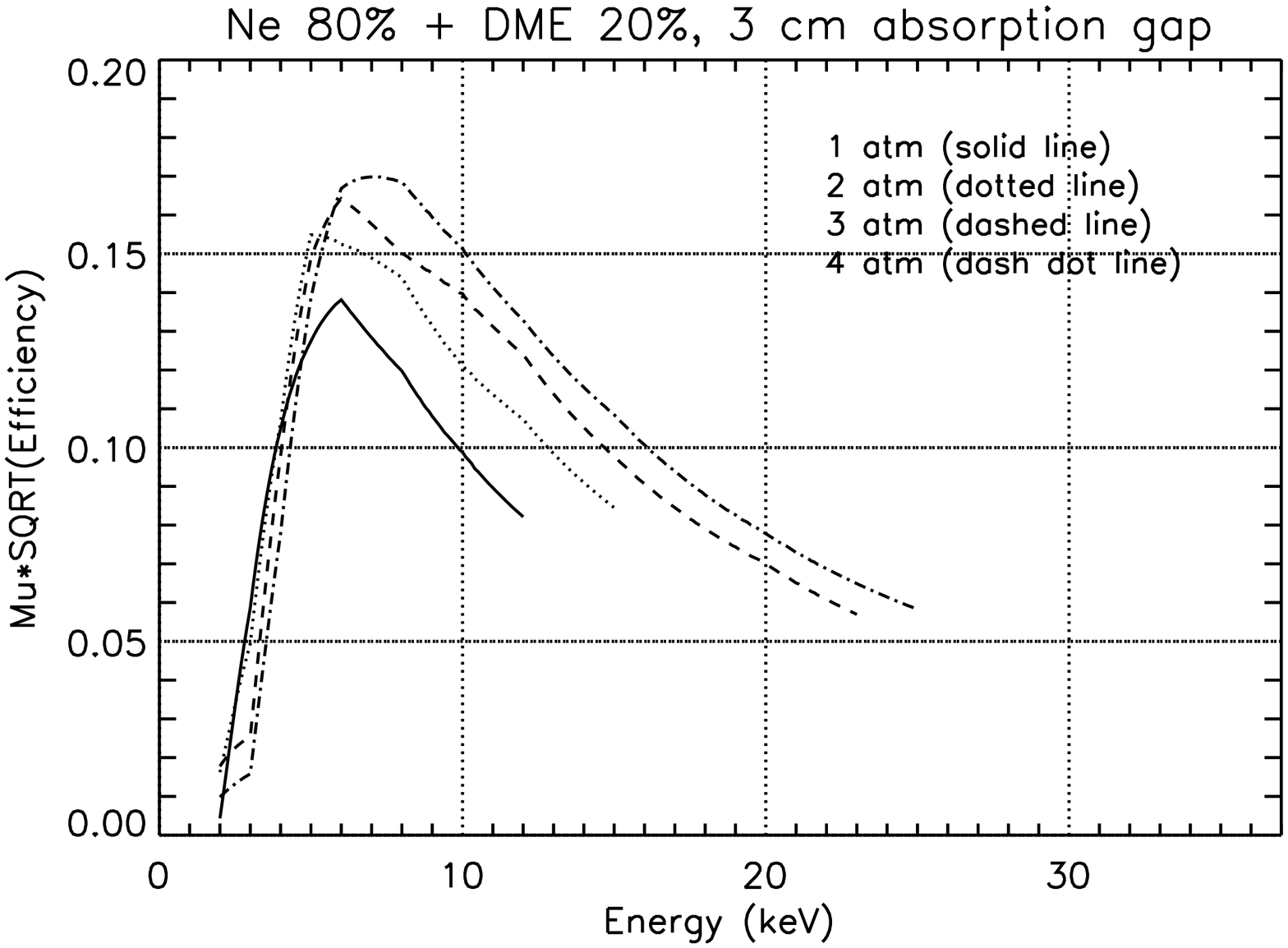}}
\end{tabular}
\end{center}
\caption[example]{Modulation ({\bf a}, {\bf c} and {\bf e}) and sensitivity factor $\mu \sqrt{\epsilon}$ ({\bf b}, {\bf d} and {\bf f}) vs photon energy for Ne~80\%~+~DME~20\% mixture with 1~cm, 2~cm and 3~cm absorption gap thickness respectively. Curves end indicates energy at which secondary charges are collected out of ASIC chip. Efficiency includes a 50~$\mu$m beryllium window.}
\label{fig:NeDME8020}
\end{figure}

In fig.~\ref{fig:Modulation_NeDME8020_1cm}, \ref{fig:Modulation_NeDME8020_2cm} and \ref{fig:Modulation_NeDME8020_3cm} we plotted modulation factor vs energy dependence for each mixture analysed. The factor $\mu \sqrt{\epsilon}$, which is an optics independent sensitivity estimate, is reported in fig.~\ref{fig:MuSqrtE_NeDME8020_1cm}, \ref{fig:MuSqrtE_NeDME8020_2cm} and \ref{fig:MuSqrtE_NeDME8020_3cm}. End of the curves indicate photon energy at which charges, generated by photoelectron and amplified from GEM, are collected out of ASIC chip.

Increasing mixture pressure and absortion gap thickness allows to reach higher peak sensitivity and larger energy band, both shifted at higher energy. For example, a 2~atm pressure mixture with an absorption gap thickness of 1~cm is sensitive above 3~keV (modulation greater than $\sim$10\%) up till 25~keV, peaking at about 5~keV, while sensitivity of a 4~atm 3~cm mixture peaks at 8~keV and is very high from 4~keV up till 25~keV.


\subsubsection{Argon Mixtures}
\label{sect:Argon}

\begin{figure}
\begin{center}
\begin{tabular}{c}
\subfigure[\label{fig:Modulation_ArDME6040_1cm}]{\includegraphics[angle=90,width=7.35cm]{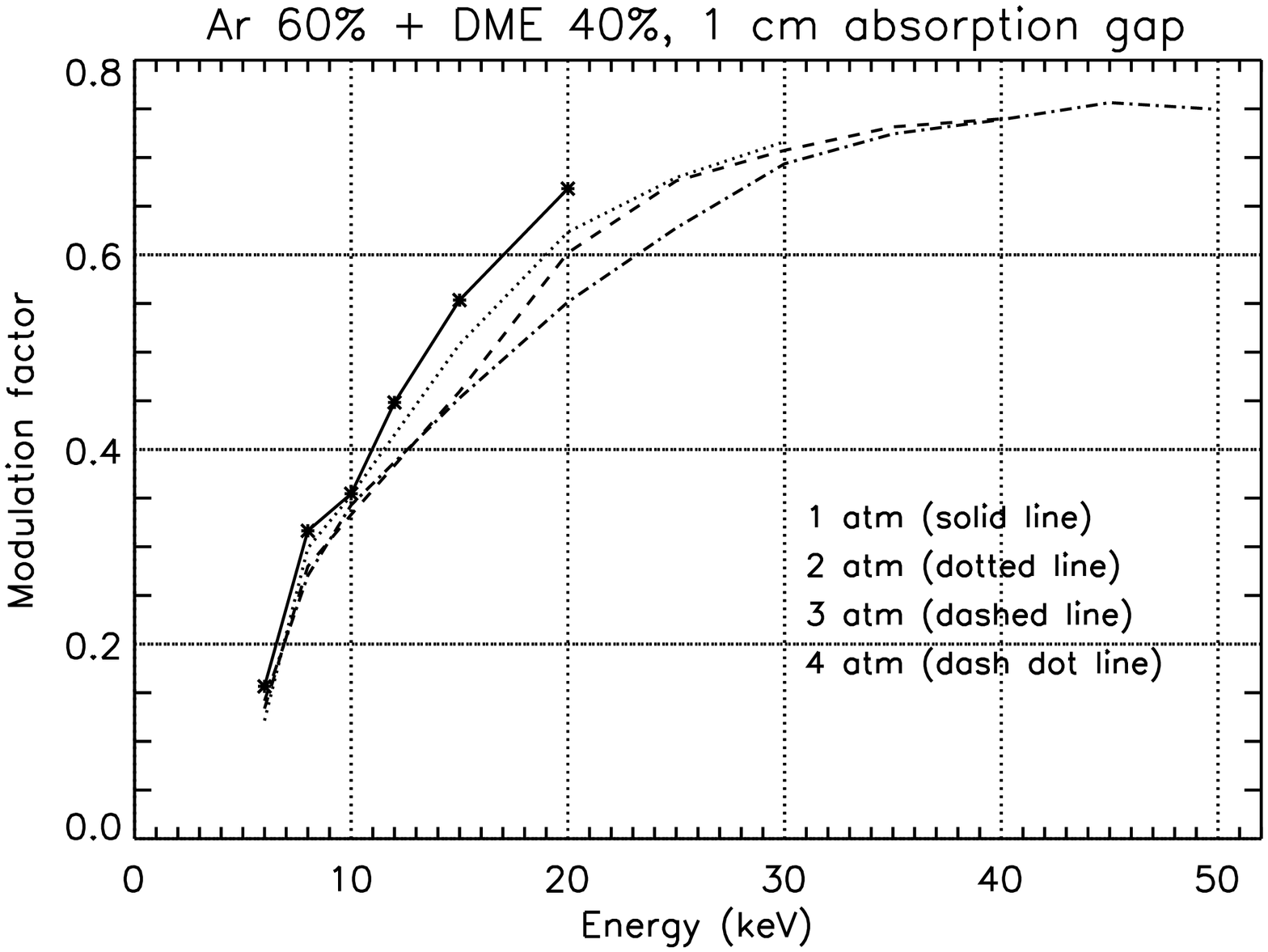}}
\subfigure[\label{fig:MuSqrtE_ArDME6040_1cm}]{\includegraphics[angle=90,width=7.35cm]{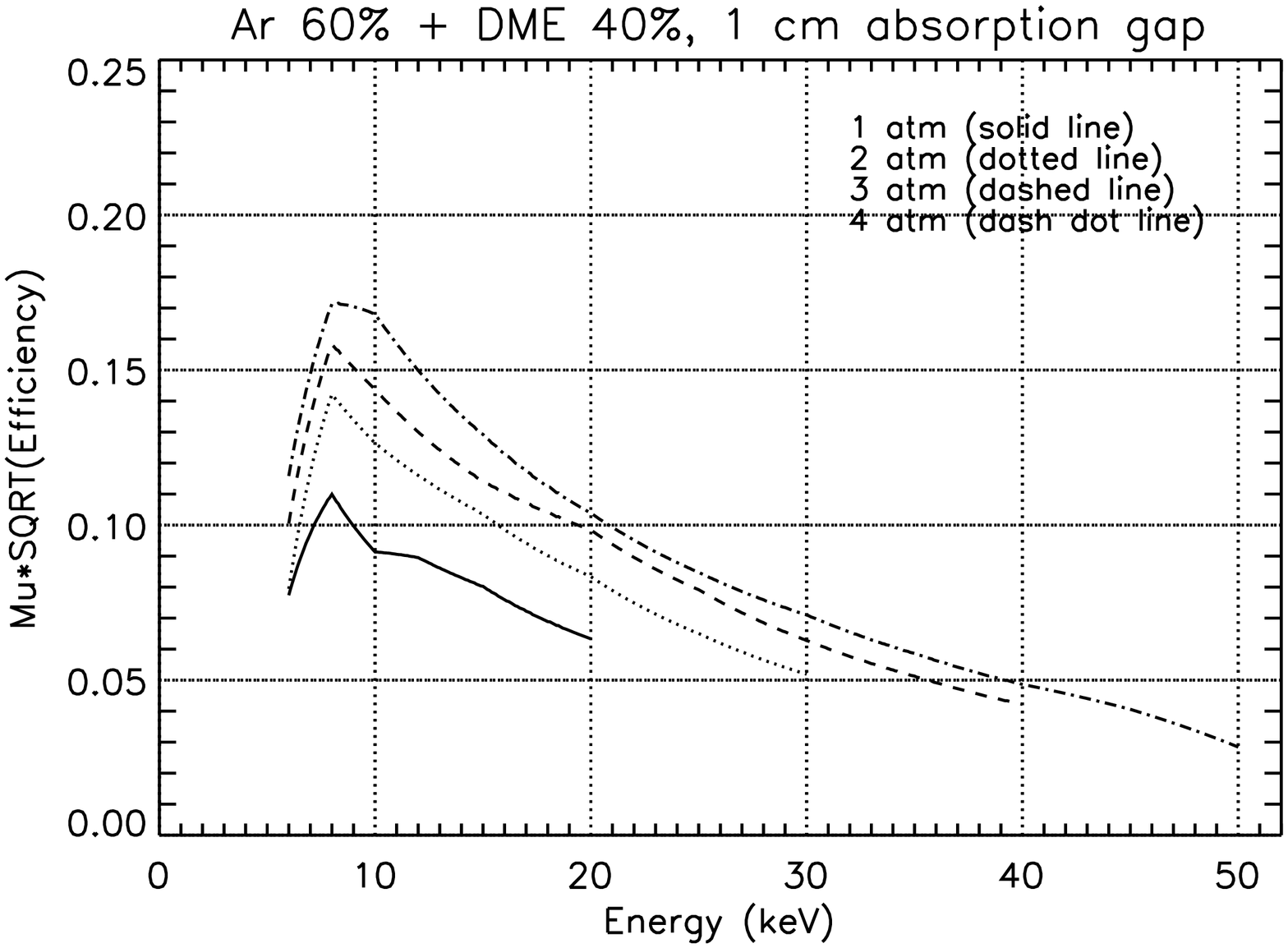}}
\\
\subfigure[\label{fig:Modulation_ArDME6040_2cm}]{\includegraphics[angle=90,width=7.35cm]{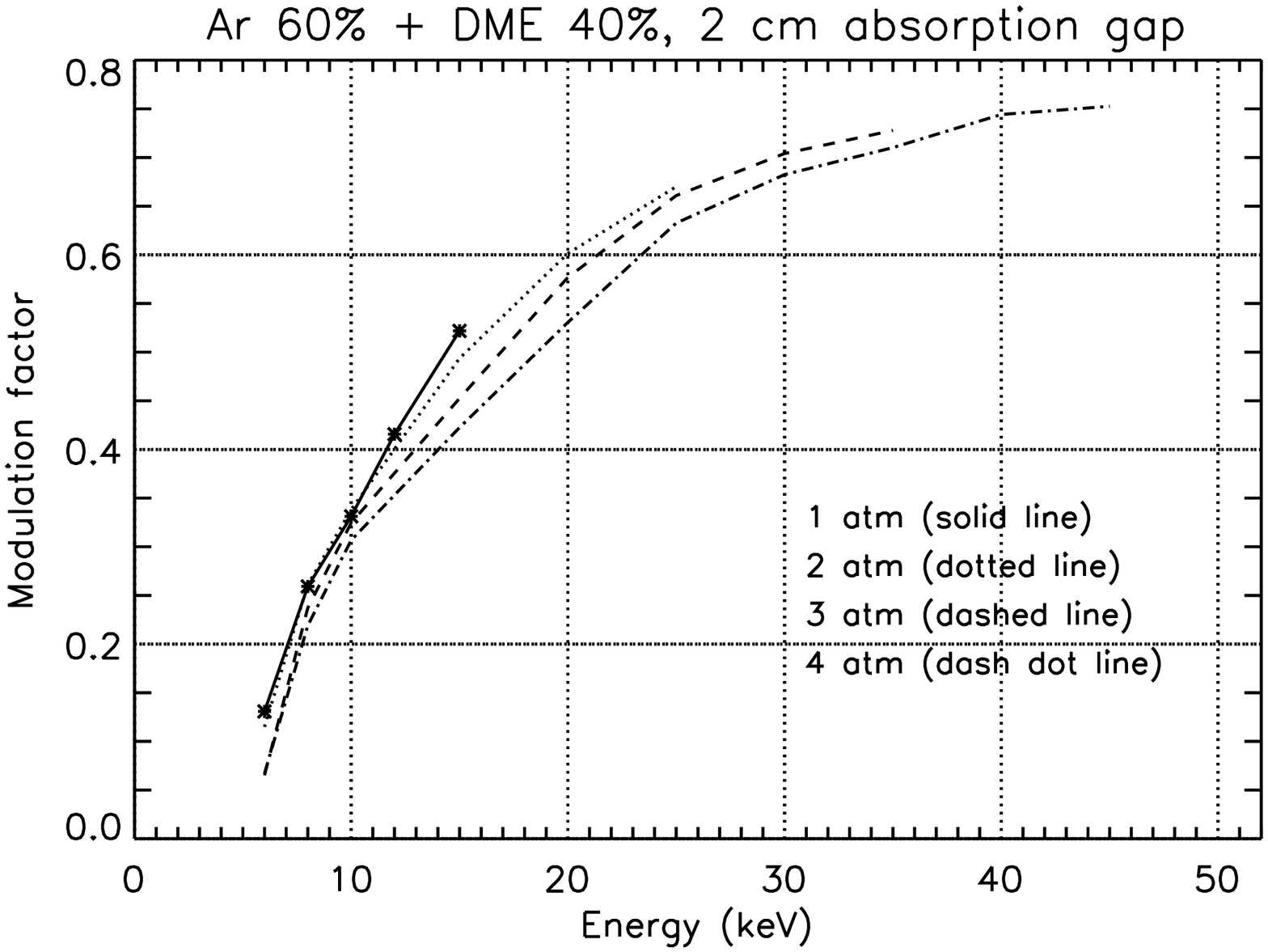}}
\subfigure[\label{fig:MuSqrtE_ArDME6040_2cm}]{\includegraphics[angle=90,width=7.35cm]{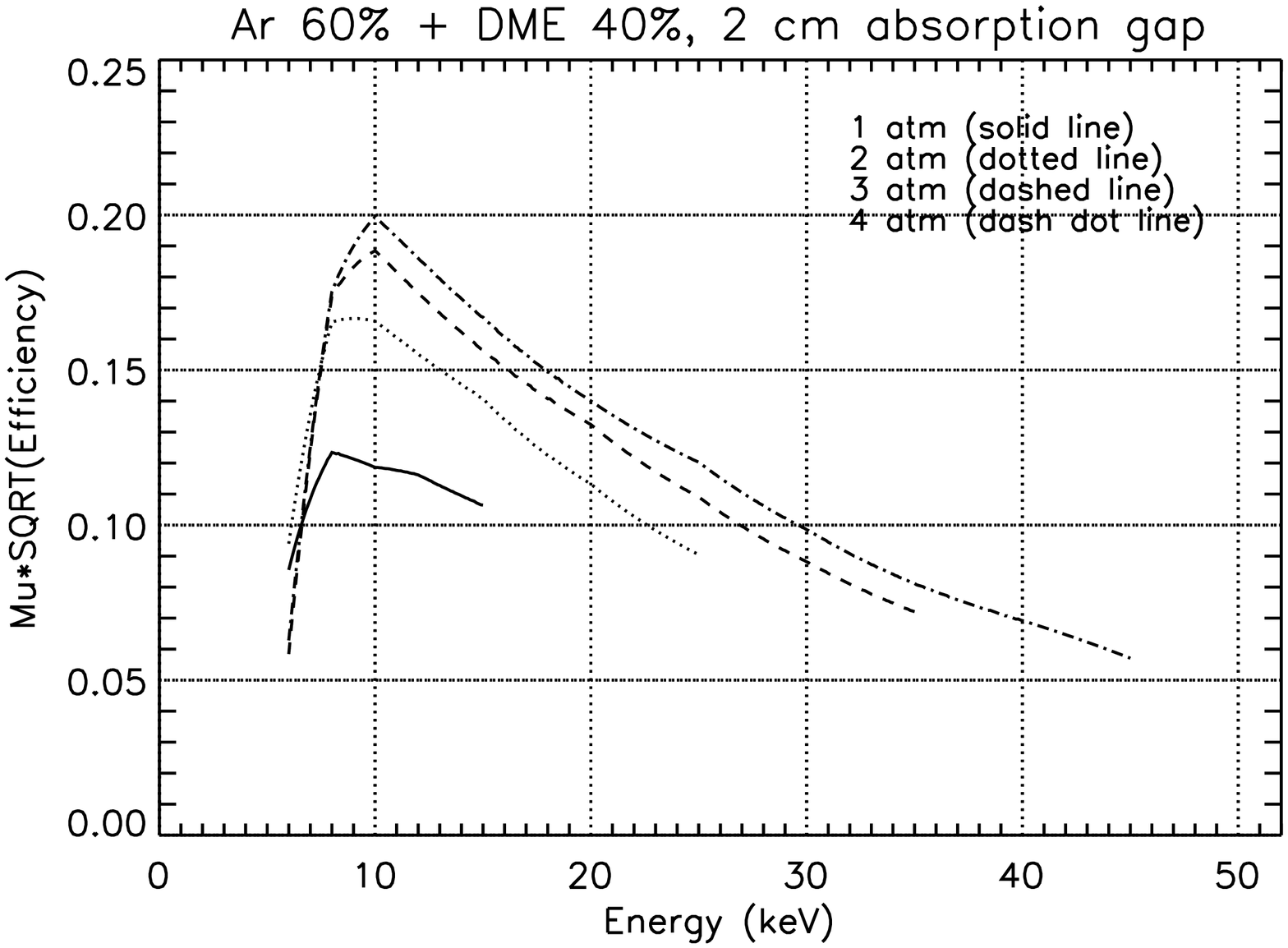}}
\\
\subfigure[\label{fig:Modulation_ArDME6040_3cm}]{\includegraphics[angle=90,width=7.35cm]{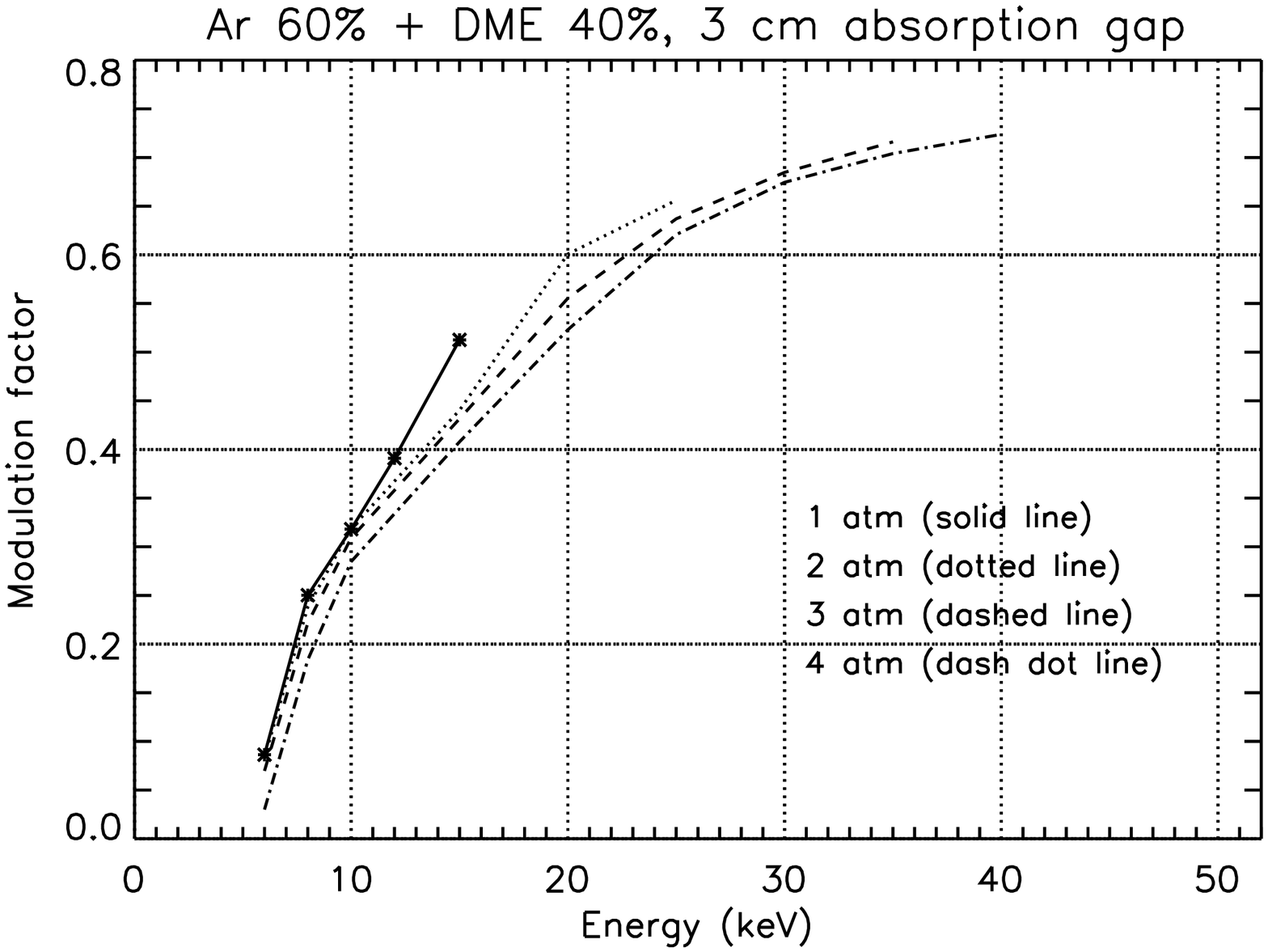}}
\subfigure[\label{fig:MuSqrtE_ArDME6040_3cm}]{\includegraphics[angle=90,width=7.35cm]{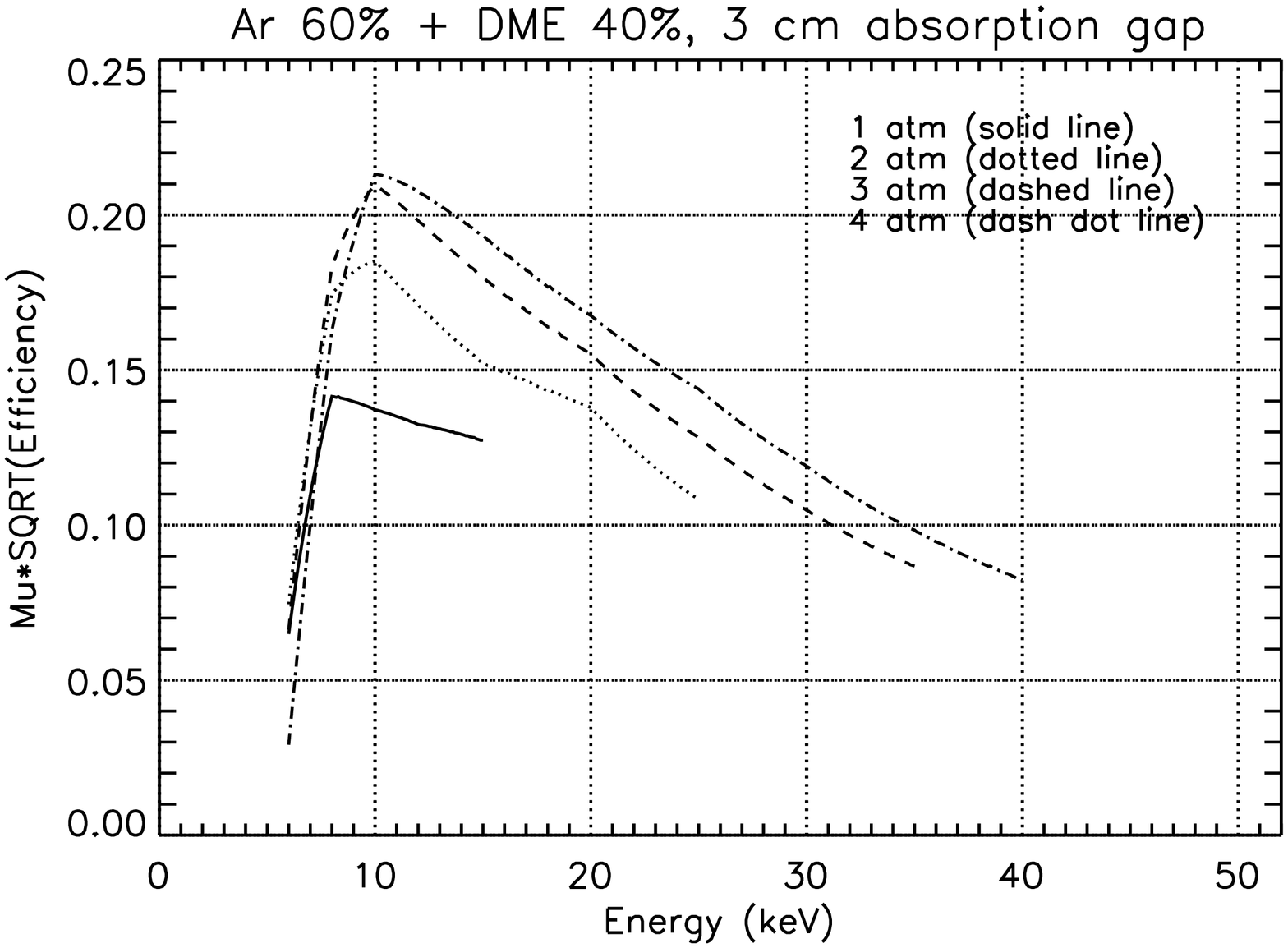}}
\end{tabular}
\end{center}
\caption[example]{Modulation ({\bf a}, {\bf c} and {\bf e}) and sensitivity factor $\mu \sqrt{\epsilon}$ ({\bf b}, {\bf d} and {\bf f}) vs photon energy for Ar~60\%~+~DME~40\% mixture with 1~cm, 2~cm and 3~cm absorption gap thickness respectively. Curves end indicate energy at which secondary charges are collected out of ASIC chip. Efficiency includes a 50~$\mu$m beryllium window.}
\label{fig:ArDME6040}
\end{figure}

We repeated sensitivity calculations for Ar~60\%~+~DME~40\% at 1~atm, 2~atm 3~atm and 4~atm pressure with 1~cm, 2~cm and 3~cm: results are reported in fig.~\ref{fig:ArDME6040}.

The modulation factor (fig.~\ref{fig:Modulation_ArDME6040_1cm}, \ref{fig:Modulation_ArDME6040_2cm} and \ref{fig:Modulation_ArDME6040_2cm}) does not decrease significantly if mixture pressure or absorption gap thickness are increased: this allows to reach high sensitivity even at very high energy without a significant modulation reduction. Sensitivity is higher than Neon based mixtures and energy band is larger, balancing the intrinsic lower fluxes from astronomical sources at high energy.

\section{Discussion}

The sensitivity curves presented in fig.~\ref{fig:NeDME8020} and \ref{fig:ArDME6040} prove that MPGC can be easly adapted to hard X-ray polarimetry. Of particular interest is the possibility of tuning the instrument to the optics: choosing Neon or Argon based mixture, a suitable pressure and absorption gap thickness, the best sensitivity can be shifted into the telescope band-pass.

A Neon based mixture at intermediate pressure and absorption gas thickness provides good modulation factor even at lower energy where sources fluxes are larger. For example, MPGC filled with a mixture of 80\% of Neon and DME at 2~atm pressure and 2~cm gap thickness provides a good sensitivity above 4~keV up till 15~keV. Polarimetry at higher energy can be performed by filling MPGC with Argon based mixtures.

A better scientific knowledge would be reached in a two telescope mission with two MPGC filled with Neon and Argon based mixtures: in this configuration X-ray polarization could be measured in different energy bands. This would provide a deeper understand of astronomical X-ray sources because emission and transport processes change radiation polarization with energy. The low energy MPGC could be filled with Neon at low pressure with an high component of DME to reduced diffusion of photoelectrons and secondary charges or with another suitable mixture (like DME + He, which is sensitive above 2~keV up till $\sim$10~keV, see ref.~\citenum{Bellazzini06c}), while the high energy specialized MPGC could be filled with an Argon based mixture at high pressure and high gap thickness. For example, sensitivity of Argon 60\% + DME 40\% mixture at 3~atm pressure and 3~cm gap thickness peaks at 12~keV and remains significant till 30~keV\footnote{Better sensitivity can be reached with mixture at 4~atm pressure, but gas discharges prevent to obtain high gain in GEM charge moltiplication.}.

\subsection{Optics integration}

We report now the performance that could be reached placing MPGC at focus of SIMBOL-X optics. We concentrate our attention on a mixture of Ar~60\% + DME~40\% at 3~atm pressure and 3~cm gap thickness, as the best choice for exploring polarimetry in hard X-rays.

\begin{figure}
\begin{center}
\begin{tabular}{c}
\includegraphics[angle=90,height=9cm]{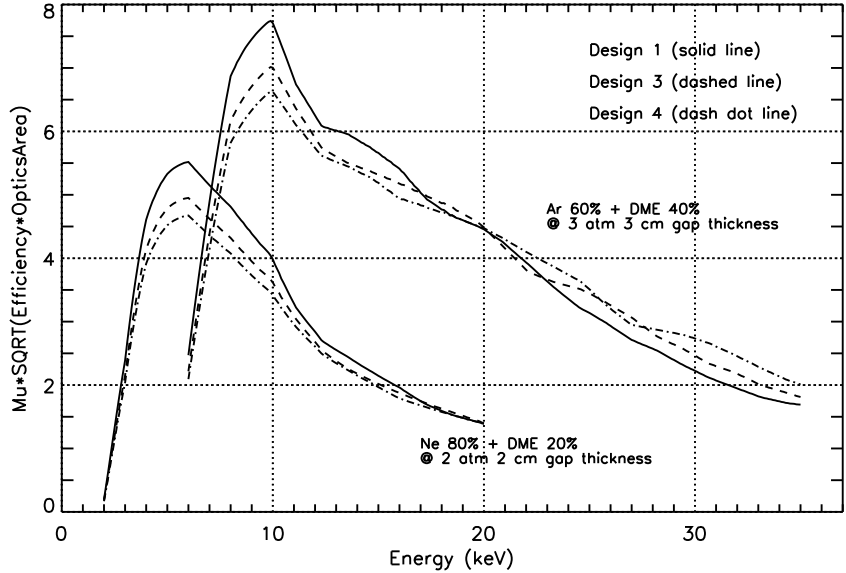}
\end{tabular}
\end{center}
\caption[example]{Sensibility factor $\mu \sqrt{\epsilon~A}$ for SIMBOL-X designs considered for Ne~80\% + DME~20\% at 2~atm pressure and 2~cm gap thickness and 3~atm 3~cm Ar~60\% + DME~40\%.}
\label{fig:MuSqrtE_NeAr}
\end{figure}

In fig.~\ref{fig:MuSqrtE_NeAr} we report $\mu \sqrt{\epsilon~A}$ factor for this mixture and three SIMBOL-X design considered: for comparison, the sensitivity of a Neon based mixture is plotted. Among three different SIMBOL-X versions analyzed, the first one is preferable because of larger area below 20~keV.

\begin{figure}
\begin{center}
\begin{tabular}{c}
\includegraphics[angle=90,width=12cm]{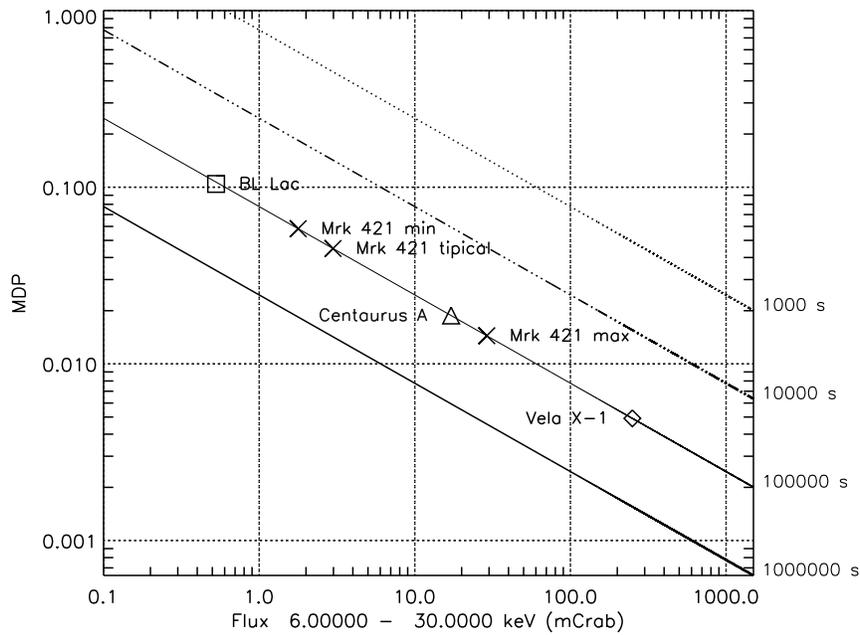}
\end{tabular}
\end{center}
\caption[example]{Minimum detectable polarization vs flux (expressed in mCrab) in the band 6-30~keV reached placing in the focus of the SIMBOL-X optics (design~1) a MPGC filled with Argon based mixture. On the right the observation time in seconds is reported: a few of representative sources are shown assuming an observation time of 10$^5$~sec.}
\label{fig:MDPvsTime_argon_6-30keV}
\end{figure}

The Argon based mixture allows a polarization measurement between 6 and 30~keV: the minimum detectable polarization reached placing in the focus of design 1 SIMBOL-X optics is plotted in fig.~\ref{fig:MDPvsTime_argon_6-30keV} for different observation times.

The high sensitivity and large energy band would allow for a spectral study of polarization in this energy range: we present now this analysis for three sources, from which high polarization is expected.

\subsubsection{Cyclotron lines in pulsating transient X0115+63 spectrum}

\begin{figure}
\begin{center}
\begin{tabular}{c}
\includegraphics[angle=0,width=7cm]{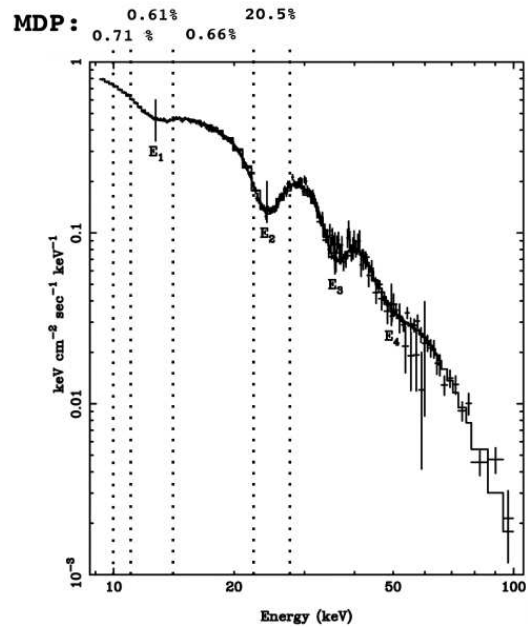}
\end{tabular}
\end{center}
\caption[example]{Minimum detectable polarization reached for pulsating transient X0115+63. Cyclotron resonant scattering feature at 12.7, 24.2, 35.7 and 49.5~keV are expected to be nearly completely polarized. Polarization measurement above 27~keV are not feasible with configuration considered. An observation time of 10$^5$~sec is assumed (adapted from ref.~\citenum{Santangelo99}).}
\label{fig:MDP_X0115}
\end{figure}

The spectrum of pulsating transient X0115+63 shows 4 cyclotron resonant scattering feature\cite{Santangelo99} at 12.7, 24.2, 35.7 and 49.5~keV, which are expected to be nearly completely polarized\cite{Meszaros88}. An Argon based MPGC could perform polarization measure till the second line: in fig.~\ref{fig:MDP_X0115} we reported expected MDP in three energy bands, 11-14~keV, 14-22~keV and 22-27~keV.

\subsubsection{Polarization measurement of Vela~X-1 pulse}

Phase and energy-dependent polarization analysis for eclipsing high mass X-ray pulsar Vela~X-1 would be of great interest to discriminate among different models which on the other hand give similar spectra: variable polarization $>$10\% is expected\cite{Meszaros88}. Basing on spectrum studies reported in ref.~\citenum{LaBarbera03}, we evaluate MDP for first minimum and first part of the ascent edge of the first pulse peak: results are reported in tab.~\ref{tab:MDP_Vela}. Thanks to the large source flux, very accurate polarization measurements are possible, allowing the detailed description of neutron star magnetosphere and the determination of spin axis and magnetic field geometry.

\begin{table}[ht]
\caption{MDP estimate of first minimum and first part of the ascent edge of the first pulse peak of Vela~X-1 spectrum with an observation time of 10$^5$~sec.}
\label{tab:MDP_Vela}
\begin{center}
\begin{tabular}{|l|c|c|} 
\hline
\rule[-1ex]{0pt}{3.5ex}   Energy band (keV)   & MDP for first minimum  & MDP for first part of the ascent edge of the first peak \\
\hline
\rule[-1ex]{0pt}{3.5ex}   6-15                & 0.0068                 & 0.0063    \\
\hline
\rule[-1ex]{0pt}{3.5ex}   15-30               & 0.0162                 & 0.0144    \\
\hline
\hline
\rule[-1ex]{0pt}{3.5ex}   6-10                & 0.0088                 & 0.0082    \\
\hline
\rule[-1ex]{0pt}{3.5ex}   10-15               & 0.0096                 & 0.0088    \\
\hline
\rule[-1ex]{0pt}{3.5ex}   15-20               & 0.0171                 & 0.0151    \\
\hline
\rule[-1ex]{0pt}{3.5ex}   20-30               & 0.0507                 & 0.0447    \\
\hline
\end{tabular}
\end{center}
\end{table}

\subsubsection{BL Lacertae objects: Markarian 501}

Markarian~501 is one of the closest (z=0.034) and brightest at all wavelengths BL Lacertae objects. Its spectrum in X-ray can be fitted with a power law of spectral index $\alpha\sim 1.5$ and can be explained with Syncroton Self Compton emission, which should be polarized at a level $\geq$10\%.\cite{Poutanen94}$^,$\cite{Celotti94} The measure of polarization can enlighten the physics of the jet: in tab.~\ref{tab:MDP_Mrk501} we reported the MDP evaluated in two energy bands for 3~atm 3~cm Argon based mixture and SIMBOL-X optics.


\begin{table}[ht]
\caption{MDP estimate for Mrk~501 at low state (power law spectral index $\sim 1.5$) for an observation time of 10$^5$~sec.}
\label{tab:MDP_Mrk501}
\begin{center}
\begin{tabular}{|l|c|c|c|} 
\hline
\rule[-1ex]{0pt}{3.5ex}   Energy band (keV)   & Min      & Max      & Outburst \\
\hline
\rule[-1ex]{0pt}{3.5ex}   6-15                & 0.0412   & 0.0174   & 0.0201   \\
\hline
\rule[-1ex]{0pt}{3.5ex}   15-30               & 0.1068   & 0.0450   & 0.0433   \\
\hline
\end{tabular}
\end{center}
\end{table}

%

\section{Conclusions}

We calculated sensitivity to hard X-ray polarization of Micropattern Gas Chamber, assuming Neon and Argon based mixures at different pressure and absoprtion gap thickness. Results demonstrate that current instruments filled with Argon + DME mixture can perform astronomical sources polarimetry up till $\sim$30~keV, allowing the study of compact galactic and extragalactic sources like X-ray pulsar or blazar. For bright sources, spectral dependence of polarization could be measured in a day of observation time, opening a new view on studies of high energy process in astrophysics.

\subsection{Acknowledgments}
This research is sponsored by INFN, INAF and ASI.

\bibliography{report}   

\begin{thebibliography}{10}

\bibitem{Pareschi03}
G.~Pareschi and V.~Cotroneo, ``Soft (0.1 - 10 kev) and hard ($>$ 10 kev) x-ray
  multilayer mirrors for the {XEUS} astronomical mission,'' in {\em SPIE
  Proc.},  ~{\bf 5168}, p.~53, 2003.

\bibitem{Pareschi04}
G.~Pareschi, V.~Cotroneo, D.~Spiga, D.~Vernani, M.~Barbera, M.~A. Artale,
  A.~Collura, S.~Varisco, G.~Grisoni, G.~Valsecchi, and B.~Negri,
  ``Astronomical soft x-ray mirrors reflectivity enhancement by multilayer
  coatings with carbon overcoating,'' in {\em SPIE Proc.},  ~{\bf 5488},
  p.~481, 2004.

\bibitem{Weisskopf76}
M.~C. {Weisskopf}, G.~G. {Cohen}, H.~L. {Kestenbaum}, K.~S. {Long},
  R.~{Novick}, and R.~S. {Wolff}, ``{Measurement of the X-ray polarization of
  the Crab Nebula},'' {\em The Astrophysical Journal}~{\bf 208},
  pp.~L125--L128, Sept.~1976.

\bibitem{Kaaret89}
P.~E. {Kaaret}, R.~{Novick}, C.~{Martin}, T.~{Hamilton}, R.~{Sunyaev}, I.~Y.
  {Lapshov}, E.~H. {Silver}, M.~C. {Weisskopf}, R.~F. {Elsner}, G.~A. {Chanan},
  G.~{Manzo}, E.~{Costa}, G.~W. {Fraser}, and G.~C. {Perola}, ``{SXRP: a focal
  plane stellar x-ray polarimeter for the Spectrum-X-Gamma mission},'' in {\em
  Proc. SPIE Vol. 1160, p. 587-0, X-Ray/EUV Optics for Astronomy and
  Microscopy, Richard B. Hoover; Ed.},  R.~B. {Hoover}, ed., pp.~587--0,
  July~1989.

\bibitem{Costa01}
E.~{Costa}, P.~{Soffitta}, R.~{Bellazzini}, A.~{Brez}, N.~{Lumb}, and
  G.~{Spandre}, ``{An efficient photoelectric X-ray polarimeter for the study
  of black holes and neutron stars},'' {\em Nature}~{\bf 411}, pp.~662--665,
  June~2001.

\bibitem{Bellazzini06a}
R.~Bellazzini, G.~Spandre, M.~Minuti, L.~Baldini, A.~Brez, F.~Cavalca,
  L.~Latronico, N.~Omodei, M.~M. Massai, C.~Sgr\'o, E.~Costa, P.~Soffitta,
  F.~Krummenacher, and R.~de~Oliveira, ``First light from a very large area
  pixel array for high-throughput x-ray polarimetry,'' in {\em SPIE Proc.},
  ~{\bf 6266}, 2006.

\bibitem{Bellazzini06b}
R.~Bellazzini, L.~Baldini, A.~Brez, F.~Cavalca, L.~Latronico, N.~Omodei, M.~M.
  Massai, M.~Minuti, M.~Razzano, C.~Sgr\'o, G.~Spandre, E.~Costa, and
  P.~Soffitta, ``Gas pixel detectors for high-sensitivity x-ray polarimetry,''
  in {\em SPIE Proc.},  ~{\bf 6266}, 2006.

\bibitem{Bellazzini03}
R.~Bellazzini, L.~Baldini, A.~Brez, E.~Costa, L.~Latronico, N.~Omodei,
  P.~Soffitta, and G.~Spandre, ``{A} photoelectric polarimeter based on a
  {M}icropattern {G}as {D}etector for {X}-ray astronomy,'' {\em Nuclear
  Instruments \& Methods in Physics Research}~{\bf 510}, pp.~176--184, 2003.

\bibitem{Yeh93}
J.~J. Yeh, {\em {A}tomic {C}alculation of {P}hotoionization {C}ross--{S}ections
  and {A}symmetry {P}arameters}, Gordon and Breach Science Publishers, 1993.

\bibitem{Krause79}
M.~O. Krause, ``Atomic radiative and radiationless yields for {K} and {L}
  shells,'' {\em Journal of Physical and Chemical Reference Data}~{\bf 8},
  p.~307, 1979.

\bibitem{Bellazzini06c}
R.~Bellazzini, L.~Baldini, F.~Bitti, A.~Brez, F.~Cavalca, L.~Latronico, M.~M.
  Massai, N.~Omodei, M.~Pinchera, C.~Sgr\'o, G.~Spandre, E.~Costa, P.~Soffitta,
  G.~D. Persio, M.~Feroci, F.~Muleri, L.~Pacciani, A.~Rubini, E.~Morelli,
  G.~Matt, and G.~C. Perola, ``A photoelectric polarimeter for {XEUS}: a new
  window in x-ray sky,'' in {\em SPIE Proc.},  ~{\bf 6266}, 2006.

\bibitem{Santangelo99}
A.~{Santangelo}, A.~{Segreto}, S.~{Giarrusso}, D.~{dal Fiume}, M.~{Orlandini},
  A.~N. {Parmar}, T.~{Oosterbroek}, T.~{Bulik}, T.~{Mihara}, S.~{Campana},
  G.~L. {Israel}, and L.~{Stella}, ``{A BEPPOSAX Study of the Pulsating
  Transient X0115+63: The First X-Ray Spectrum with Four Cyclotron Harmonic
  Features},'' {\em The Astrophysical Journal}~{\bf 523}, pp.~L85--L88,
  Sept.~1999.

\bibitem{Meszaros88}
{{Meszaros}, P. and {Novick}, R. and {Szentgyorgyi}, A. and {Chanan}, G.~A. and
  {Weisskopf}, M.~C.}, ``{Astrophysical implications and observational
  prospects of X-ray polarimetry},'' {\em {The Astrophysical Journal}}~{\bf
  324}, pp.~{1056--1067}, jan~1988.

\bibitem{LaBarbera03}
M.~O. A.~La~Barbera, A.~Santangelo and A.~Segreto, ``A pulse phase-dependent
  spectroscopic study of vela x-1 in the 8-100~kev band,'' {\em Astronomy {\&}
  Astrophysics}~{\bf 400}, pp.~993--1005, 2003.

\bibitem{Poutanen94}
J.~{Poutanen}, ``{Relativistic jets in blazars: Polarization of radiation},''
  {\em The Astrophysical Journal}~{\bf 92}, pp.~607--609, June~1994.

\bibitem{Celotti94}
A.~{Celotti} and G.~{Matt}, ``{Polarization Properties of Synchrotron
  Self-Compton Emission},'' {\em Monthly Notices of Royal Astronomical
  Society}~{\bf 268}, p.~451, May~1994.

\end{thebibliography}
\bibliographystyle{spiebib}   

\end{document}